          [2001/04/25 1.1 (PWD)]
\documentclass[a4paper,twocolumn]{esapub} 
\usepackage{natbib}
\usepackage{epsfig}

\newcommand{\figwidth}{0.48\textwidth}

\title{A  Chandra Deep X-ray Exposure on the Galactic Plane and  Near Infrared
 Identification}
\author{K. Ebisawa, A. Paizis, T. J.-L. Couvoisier, P. Dubath}
\affil{INTEGRAL Science Data Centre, Chemin d'\'Ecogia 16, 1290 Versoix, Switzerland}
\author{M. Tsujimoto}
\affil{Department of Astronomy and Astrophysics, Pensylvania State Univesrity, University Park,  PA 16802, USA}
\author{K. Hamaguchi, \\V. Beckmann}
\affil{Laboratory for High Energy Astrophysics, NASA/GSFC, Greenbelt, MD 20771, USA}
\author{A. Bamba, A. Senda, M. Ueno}
\affil{Department of Physics, Kyoto University, Kitashirakawa Oiwake-cho, Sakyo-ku, Kyoto, 606-8502, Japan}
\author{H. Kaneda, Y. Maeda, G. Sato}
\affil{Institute of Space and Astronautical Science, Yoshinodai, Sagamihara, Kanagawa, 229-8510 Japan}
\author{S. Yamauchi}
\affil{Faculty of Humanities and Social Sciences, Iwate University, Iwate, 020-8550, Japan}
\author{R. Cutri}
\affil{IPAC, California Institute of Technology, code 100-22, 770 South Wilson Avenue, Pasadena, CA 91125, USA} 
\author{E. Nishihara}
\affil{Gunma Astronomical Observatory,  Nakayama Takayama-mura, Agatsuma-gun Gunma, 377-0702, Japan}

\begin{document}


\maketitle

\begin{abstract}
Using the {\it Chandra} ACIS-I instruments, we have carried out a deep X-ray observation on the Galactic plane 
region at  $(l,b) \approx (28.^\circ5,  0.^\circ0)$, where no discrete X-ray sources have been
known previously.  We have detected, as well as strong diffuse emission, 
274 new point X-ray sources (4 $\sigma$ confidence) 
within two partially overlapping fields ($\sim $250 arcmin$^2$ in total)
down to the flux limit  $\sim 3 \times 10^{-15} $ 
erg s$^{-1}$ cm$^{-2}$
(2 -- 10 keV) 
and   $\sim 7 \times 10^{-16} $ erg s$^{-1}$ cm$^{-2}$ (0.5 -- 2 keV). 
We clearly resolved point sources and the Galactic diffuse emission, and
 found that $\sim 90$ \%  of the flux observed in
our field of view originates from  diffuse emission.
Many point sources are detected {\em either}\/ in the soft X-ray band (below 2 keV) or in the
hard band (above 2 keV), and
only a small number of sources are detected in both energy bands.
 On the other hand, most soft X-ray sources are considered to be
nearby X-ray active stars.  
We have carried out a follow-up
near-infrared (NIR) observation using SOFI at ESO/NTT.  Most of the soft X-ray sources
were identified,  whereas only a small number of  hard X-ray sources had counterparts in NIR.
Using both X-ray and NIR
information, we can efficiently classify the point X-ray sources detected in the
Galactic plane. We conclude that most of the hard X-ray sources are background Active Galactic Nuclei seen through
the Milky Way, whereas majority of the soft X-ray sources are nearby X-ray active stars.
\end{abstract}

\begin{figure}[t]
\centerline{
\epsfig{figure=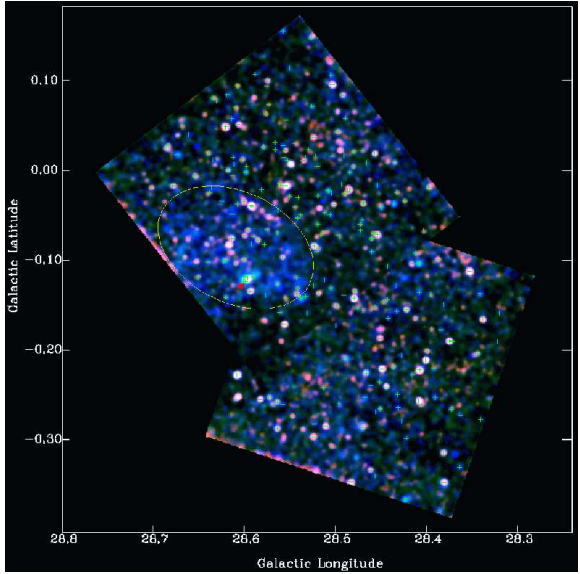,angle=0,width=7.6cm}
}
\caption{
Superimposed image of the  two {\it Chandra} observations (exposure and
vignetting are corrected) in Galactic coordinates.
South is AO1  and North is the AO2 field, each for 100 ksec exposure. This is a pseudo-color
image such that soft X-rays in 0.5 -- 2 keV, 
medium X-rays in 2 -- 4 keV and  hard hard X-rays in 4 -- 8 keV are
represented in red, green and blue, respectively (in the electronic version).  
The image is adoptively smoothed so that both the point sources and the
diffuse emission are clearly  visible.
The 274 detected point source are marked with crosses.
The region including the supernova remnant AX J 1843.8--0352/G28.6--0.1
(Bamba et al.\ 2001; Ueno et al.\ 2003) is shown with the yellow ellipse,
within which an extended thermal blob named 
CXO J184357-35441 (Ueno et al.\ 2003) is marked with the red arrow.
Note that the supernova remnant AX J 1843.8--0352/G28.6--0.1 is prominent
in hard X-rays (``bluish'' in this representation).
}
\end{figure}

\section{Chandra observation and results}
We have carried out two 100 ksec pointings with {\it Chandra} ACIS-I
in AO1 (February 25 and 26, 2000) and AO2
(May 20, 2001),  with slightly overlapping 
fields (Fig.\ 1).  Total area of the observed field is $\sim 250 $ arcmin$^2$. The first results from the AO1 observation 
has been published in Ebisawa et al.\ (2001).
Detailed analysis of the new supernova remnant candidate in the field of view, AX J1843.8--0352/G28.6--0.1 (marked in
Fig.\ ),   is reported in Ueno et al.\ (2003).
The full results including the X-ray source catalog will be published in Ebisawa et al.\ (2004).


We have extracted an energy spectrum from  our 
{\it Chandra} Galactic plane field (Fig.\ 2) by subtracting the instrumental background
and excluding the AX J 1843.8--0352/G28.6--0.1 region (marked in Fig.\  1).
We found that the diffuse emission contributes to 
$\sim 90$ \% of the flux in the field of view, and that emission lines
from highly ionized heavy elements are associated with the diffuse emission.
This indicates that the Galactic ridge X-ray emission is truly diffuse, and its
origin is probably a highly ionized plasma.


We have made $\log N- \log S$ curves  in 2 -- 10 keV and 
0.5 -- 2 keV from the surface density of the detected point sources
(Fig.\  3).
Our results are also compared with those from the high Galactic region and 
Galactic center.

In the 2 -- 10 keV band, after taking into account  the Galactic absorption,
our $\log N- \log S$ curve does not indicate significant excess over the
extragalactic one.  This indicates that most of the hard X-ray point sources
we detected in the Galactic plane are in fact of extragalactic origin, presumably
background AGNs.  If we compare the hard X-ray  $\log N-\log S$ curve
at the Galactic center region and that of the Galactic plane (Fig.\ 3, left), we see clear 
excess of the Galactic X-ray sources in the Galactic center region over the
Galactic plane.
In the soft X-ray band, on the other hand,
contribution from the extragalactic sources is negligible since they
are almost fully absorbed.  Thus almost  all the soft X-ray sources discovered
in our Galactic plane field are considered to be Galactic.

\section{Near Infrared Observation at ESO using NTT/SOFI}

Because of the heavy Galactic absorption, the near infrared band has 
a great advantage over the optical  to identify  dim X-ray
sources in the Galactic plane.
In order to identify  X-ray point sources in our {\it Chandra} field, 
we  have carried out a NIR follow-up observation 
at European Southern Observatory (ESO) using the New Technology Telescope 
with the SOFI  infrared camera.  The observations were
carried out on the nights of July 28 and 29, 2002.

\begin{figure}[t]
\centerline{
\epsfig{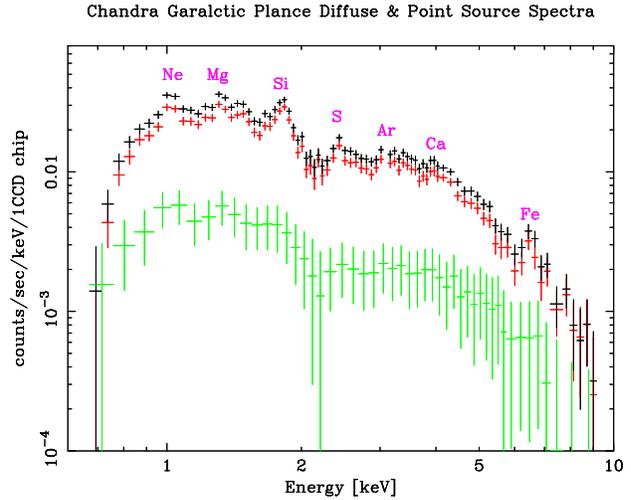}
}
\caption{
X-ray energy spectrum of our {\it Chandra} Galactic plane field.
The AX J 1843.8--0352/G28.6--0.1 region (marked in Fig.\   1)
was excluded.  Black (top) is the total emission in the field
of view, and green (bottom) is the sum of the point sources, red
(slightly lower than the black) is the residual diffuse emission. 
}
\end{figure}

In Fig.\  4, we plot angular distance between 
the {\it Chandra} point sources (classified with X-ray colors) and the nearest 
NIR sources detected by SOFI.
If the separation between the {\it Chandra} source and SOFI source is less than $\sim 1.0''$, we identify
the two sources.  We can see that most of the soft X-ray 
sources (marked in red) have NIR counterparts, while only a small
portion of the hard X-ray sources is  identified in the NIR.
This is consistent with the result from the $\log N-\log S$
analysis that most hard X-ray sources are extragalactic,
whereas almost all the soft X-ray sources are Galactic.

\begin{figure*}[htbp]
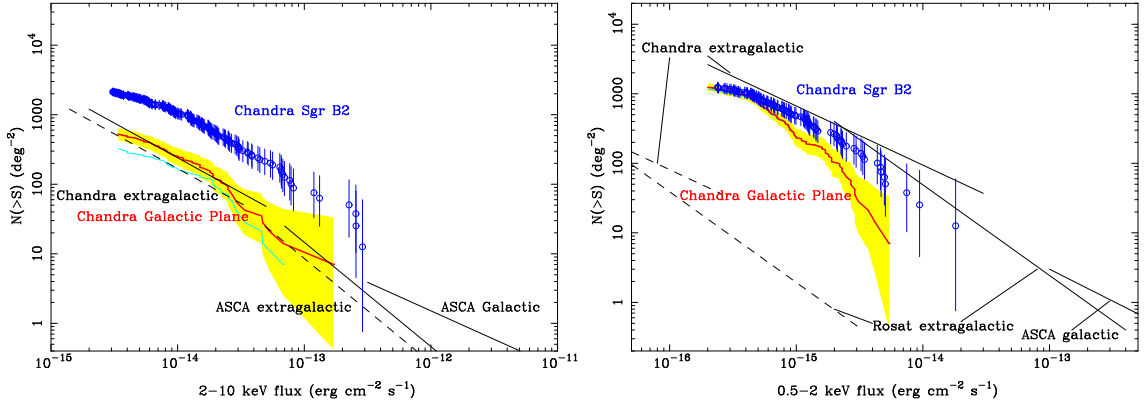

\centerline{
\epsfig{figure=f3-1.eps,angle=270,width=7.6cm}
\epsfig{figure=f3-2.eps,angle=270,width=7.6cm}
}
\caption{The $\log N - \log S$ curves of the point sources detected in  our {\it Chandra} Galactic
plane field in 2 -- 10 keV (left) and 0.5 -- 2 keV (right). They are indicated in red lines
(in the electronic version) within the 90 \% error regions (polygon shape in yellow).  
Together, other  $\log N - \log S$
relations are shown for the bright {\it ASCA} Galactic sources (Sugizaki et al.\  2001), {\it Chandra} Galactic center (Sgr B2) point sources
and extragalactic point sources detected with {\it ASCA} (Ueda et al.\ 1999), {\em ROSAT} and {\it Chandra} 
(Giacconi et al.\ 2001).  For the extragalactic sources, both the original $\log N - \log S$
curves at high Galactic latitudes  and the ones expected on the Galactic plane 
with a hydrogen column density of $6 \times 10^{22}$ cm$^{-2}$ (broken lines) are shown. 
Also the $\log N - \log S$ curves of  only the sources having the near infrared counterparts 
are shown  in the left figure
(slightly below our Chandra Galactic $\log N - \log S$ curve; cyan in the electronic version).
}\label{logN-logS}
\end{figure*}

\section {Classification of the Point Sources}

Using both X-ray and NIR information, we may effectively classify X-ray
point sources.

In Fig.\  5, we show a  histogram of the X-ray hardness ratio distribution 
and the intensity-hardness  diagram.  From the left panel
in Fig.\ 5, we can see that the softest sources are most numerous,
and most of them have NIR counterparts.  With increasing hardness ratio,
the number of sources first decreases, but again increases toward the
hardest values.  This clearly indicates the dichotomy of the point source population,
Galactic population and  extragalactic one.

In the right panel of Fig.\  5, we can see that all the {\em bright 
soft}\/ sources have NIR counterparts.  Only several {\em dim soft}
sources do not have counterparts, presumably because they are intrinsically
dim and/or very far Galactic sources. On the other hand, we can see that
there are many {\em bright hard}\/  sources  which 
are not identified in NIR. In particular, the brightest {\it Chandra} 
source, which has the hardness ratio 0.74 (Fig.\   5), is not identified in the NIR.
This is not surprising since AGNs tend to be bright
in hard X-rays, but even the brightest AGNs should be completely absorbed
through  the Galactic plane.

\begin{figure}[t]
\centerline{
\epsfig{figure=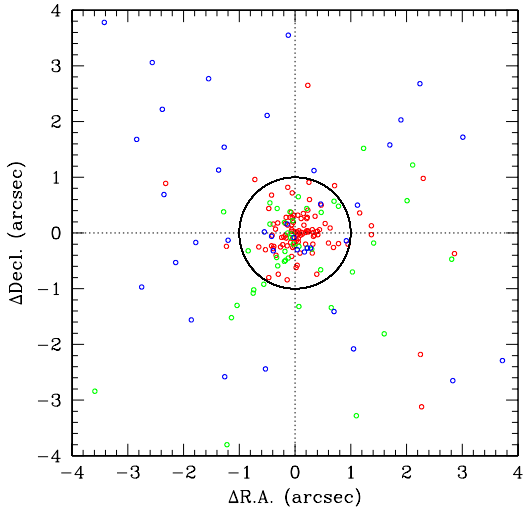,width=0.44\textwidth}
}
\caption{
Distance between the {\it Chandra} point source and the nearest
NIR source detected by SOFI. In the electronic version, 
soft, medium and hard  X-ray sources are distinguished with red, 
green and blue, respectively.
}
\end{figure}

\begin{figure*}[htbp]
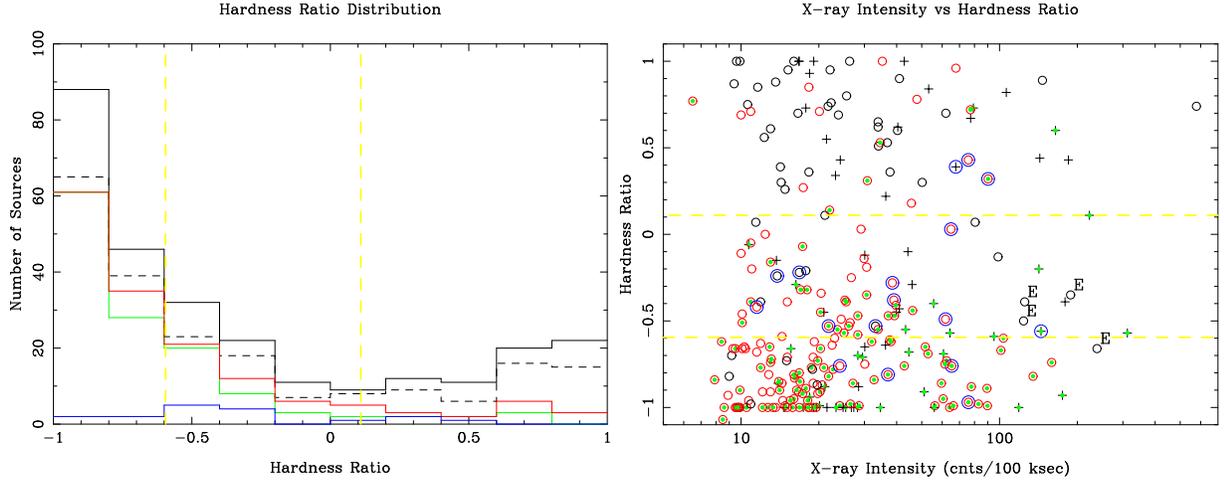

\centerline{
\epsfig{figure=f5-1.eps,angle=270,width=0.47\textwidth}
\epsfig{figure=f5-2.eps,angle=270,width=0.47\textwidth}
}
\caption{Histogram of the hardness ratio distribution 
(left) and intensity-hardness diagram (right). 
The hardness ratio is defined as $(H-S)/(H+S)$, where
$S$ and $H$ are vignetting-corrected counting rates in 
0.5 -- 2 keV and 3 -- 8 keV, respectively.
In the
left panel, the broken line indicates the number of sources within the
SOFI fields, and the line below (red in the electronic version) 
shows the sources  having the SOFI near-infrared
counterparts.  The green line (below red) tells the number of sources having
the 2MASS counterparts, and the blue line (bottom) is for the variable 
sources. In the right panel, the sources outside 
of the SOFI field are shown with crosses, and those inside are with circles. 
In the electronic version, 
black circles indicate sources without SOFI counterparts, while red circles are
those having  the SOFI counterparts.  In addition, sources having the 2MASS 
counterparts  are marked with green dots, and variable sources
are marked with blue circles. The horizontal broken lines (yellow) in both figures
indicate the boundaries  we defined between the soft and medium sources (hardness ratio=-0.595), and
the medium and hard sources (hardness ratio=0.11).
}\label{hardness}  
\end{figure*}

We have created composite (average) energy spectra of point sources,
by dividing them into four categories, depending on the X-ray hardness
(soft and hard) and presence or absence of the NIR counterparts
(Fig.\  6).  The soft sources with NIR counter parts (top left)
show thermal emission lines, and are fitted with the two-component thermal
plasma model.  This kind of spectrum is consistent with stellar coronal emission. The
soft sources without NIR counterparts are dimmer but can be fitted with
exactly the same model with decreasing  normalization and increasing
 hydrogen column density, which suggests they are farther and/or
dimmer active stars.

Hard X-ray sources with NIR counterparts, even though only handful,
indicate a clear  narrow iron emission line at 6.67 keV from helium-like ions 
and a flat continuum spectrum.  This is a signature of a high temperature thermal
plasma, expected from quiescent cataclysmic variables.  On the other hand,
the hard X-ray sources without NIR counterparts are {\it brighter}, and
they show rather complex iron features which look like broad lines or
edges.  The X-ray  brightness and complex iron feature supports the
idea that these hard X-ray sources without NIR counterparts are background AGNs.

\begin{figure*}[htbp]
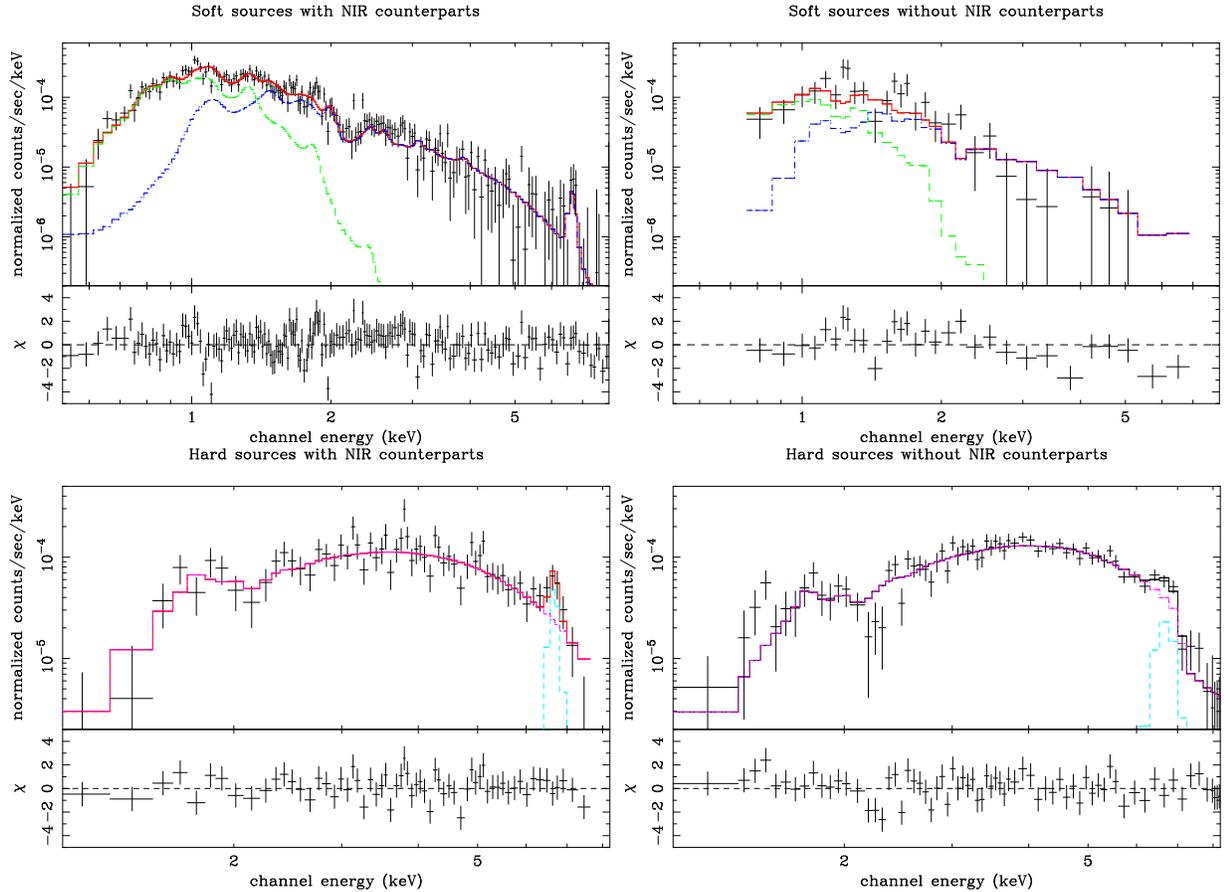

\centerline{
\epsfig{figure=soft_with_nir.ps,angle=270,width=8cm}
\epsfig{figure=soft_without_nir.ps,angle=270,width=8cm}
}
\centerline{
\epsfig{figure=nir_exist_hard.ps,angle=270,width=8cm}
\epsfig{figure=nir_does_not_exist_hard.ps,angle=270,width=8cm}
}
\caption{Composite energy spectra and model fitting 
of the point sources grouped by the X-ray spectral hardness and absence or presence of the NIR counterparts.  Those having the NIR counterparts are in 
the left-hand side, and those without the NIR counterparts are in the right-hand
side.  The top two panels are the soft source spectra, and the bottom ones the hard source spectra. 
}\label{Souce_Spectra}  
\end{figure*}


\begin{thebibliography}{}
\bibitem[]{}Bamba, A., Ueno, M., Koyama, K. Yamauchi, S. 2001, PASJ, 53, L21
\bibitem[]{}Ebisawa, K., Maeda, Y.,  Kaneda, H. and  Yamauchi, S.  2001, Science, 293, 1633
\bibitem[]{}Ebisawa, K. et al.\ 2004, in preparation
\bibitem[]{}Giacconi, R. et al.\ 2001,  ApJ, 551, 624
\bibitem[]{}Sugizaki, M. et al.\  2001, ApJS, 134, 77
\bibitem[]{}Ueda, Y. et al.\ 1999, ApJ, 518, 656 
\bibitem[]{}Ueno, M., Bamba, A., Koyama, K. and  Ebisawa, K. 2003, ApJ, 588, 338
\end{thebibliography}
\end{document}